\documentclass[iop]{emulateapj}

\usepackage{graphics}
\usepackage{epsfig}
\usepackage{natbib}
\newcommand{\kms}{km~s\ensuremath{^{-1}}}
\newcommand{\msun}{$M_{\odot}$}
\newcommand{\nuvr}{NUV$-r$}
\newcommand{\mh}{$M_{H_2}$}

\newcommand{\mstar}{$M_{\ast}$}
\newcommand{\must}{$\mu_{\ast}$}

\newcommand{\tdep}{$t_{dep}({\rm H_2})$}
\newcommand{\tdepHI}{$t_{dep}({\rm HI})$}
\newcommand{\fgas}{$f_{H_2}$}
\newcommand{\xco}{$\alpha_{CO}$}
\newcommand{\ntot}{365}

\shorttitle{Star formation efficiency in massive galaxies}
\shortauthors{Saintonge et al.}

\begin{document}

\title{The impact of interactions, bars, bulges, and AGN on star formation efficiency \\ in local massive galaxies }

\author{Am\'{e}lie Saintonge\altaffilmark{1,2}, Linda J. Tacconi\altaffilmark{2}, Silvia Fabello\altaffilmark{1},  Jing Wang\altaffilmark{1},  Barbara Catinella\altaffilmark{1}, \\
 Reinhard Genzel\altaffilmark{2},  Javier Graci\'{a}-Carpio\altaffilmark{2}, Carsten Kramer\altaffilmark{3},  Sean Moran\altaffilmark{4}, Timothy M. Heckman\altaffilmark{4}, \\
  David Schiminovich\altaffilmark{5}, Karl Schuster\altaffilmark{6}, Stijn Wuyts\altaffilmark{2}}

\altaffiltext{1}{Max-Planck Institut f\"ur Astrophysik, 85741 Garching, Germany}
\altaffiltext{2}{Max-Planck Institut f\"ur extraterrestrische Physik, 85741 Garching, Germany}
\altaffiltext{3}{Instituto Radioastronom\'{i}a Milim\'{e}trica, Av. Divina Pastora 7, Nucleo Central, 18012 Granada, Spain}
\altaffiltext{4}{Johns Hopkins University, Baltimore, Maryland 21218, USA}
\altaffiltext{5}{Department of Astronomy, Columbia University, New York, NY 10027, USA}
\altaffiltext{6}{Institut de Radioastronomie Millim\'{e}trique, 300 Rue de la piscine, 38406 St Martin d'H\`{e}res, France}

\begin{abstract}
Using atomic and molecular gas observations from the GASS and COLD GASS surveys and complementary optical/UV data from SDSS and {\it GALEX}, we investigate the nature of the variations in the molecular gas depletion time observed across the local massive galaxy population.  The large and unbiased COLD GASS sample allows us for the first time to statistically assess the relative importance of galaxy interactions, bar instabilities, morphologies and the presence of AGN in regulating star formation efficiency.   We find that both the H$_2$ mass fraction and depletion time vary as a function of the distance of a galaxy from the main sequence traced by star-forming galaxies in the SFR-\mstar\ plane. The longest gas depletion times are found in below-main sequence bulge-dominated galaxies (\must$>5\times10^8$\msun\ kpc$^{-2}$, $C>2.6$) that are either gas-poor (\mh/\mstar$<$1.5\%), or else on average less efficient by a factor of $\sim2$ than disk-dominated galaxy at converting into stars any cold gas they may have.  We find no link between the presence of AGN and these long depletion times. In the regime where galaxies are disc-dominated and gas-rich, the galaxies undergoing mergers or showing signs of morphological disruptions have the shortest molecular gas depletion times, while those hosting strong stellar bars have only marginally higher global star formation efficiencies as compared to matched control samples.   Our interpretation is that the molecular gas depletion time variations are caused by changes in the ratio between the gas mass traced by the CO(1-0) observations, and the gas mass in high density star-forming cores (as traced by observations of e.g. HCN(1-0)).  While interactions, mergers and bar instabilities can locally increase pressure and raise the ratio of efficiently star-forming gas to CO-detected gas (therefore lowering the CO-based depletion time), massive bulges may prevent the formation of dense clumps by stabilizing gas disks against fragmentation, therefore producing the long depletion times.   Building a sample representative of the local galaxy population with \mstar$>10^{10}$\msun, we derive a global Kennicutt-Schmidt star formation relation of slope $1.18\pm0.24$, and observe structure within the scatter around this relation, with galaxies having low (high) stellar mass surface densities lying systematically above (below) the mean relation, suggesting that $\Sigma_{\rm H_2}$ is not the only parameter driving the global star formation ability of a galaxy.  
\end{abstract}


\section{Introduction}
\label{intro}

Star formation in galaxies proceeds at maximum efficiency in systems undergoing major mergers \citep{genzel10,daddi10}.  Given the well-established hierarchical structure formation paradigm, where larger structures form from the merging of smaller objects, it is tempting to infer that this ``merging mode" of star formation contributes significantly to the build-up of stellar mass across the galaxy population, especially at high redshift.  However, recent observations downplay the importance of violent events such as merger-induced starbursts in the global star formation budget of the universe, even at $z\sim2$ \citep{rodighiero11}.  

Instead, the existence of a tight relation between stellar mass and star formation rate, followed by star-forming galaxies up to at least $z=2$ \citep[e.g.][]{noeske07,salim07,rodighiero10,nordon10,elbaz11}, suggests that star formation is mainly regulated by secular processes with much longer duty cycles.  To make further progress in understanding the histories of galaxies, we therefore need to investigate how processes such as minor mergers, bar formation and bulge growth influence the star formation process, both in terms of its onset/shutdown and of its efficiency. 

The first question to answer is thus, how do these mechanisms contribute to switching star formation on/off in galaxies?  Several studies have recently tackled this specific question, by looking for the observable that correlates best with quiescence.  Both observations and simulations suggest that quiescence is not strongly linked with stellar mass,  but rather with structural and morphological parameters such as Sersic index, stellar mass surface density, or central velocity dispersion \citep{kauffmann03,schiminovich07,more11,wuyts11,GASS1,GASS6,COLDGASS1,COLDGASS3}.  While there is mounting evidence for a link between star formation quenching and the presence of bulges, the direction of causality between them is as of yet not fully established.  For example, \citet{bell12} suggest that while a bulge-dominated morphology is necessary for quenching to occur, it is not sufficient in itself.  The exact underlying physical mechanism(s) responsible for the correlation between early-type morphologies and the absence of star formation has also yet to be unambiguously identified.  We come back to this discussion in \S \ref{bulges}.

After having determined which galaxies are star-forming and which are quiescent, at least in a statistical way, the second question concerns the efficiency of star formation, across the galaxy population.  The general picture is that a star-forming galaxy will convert a given amount of molecular gas into stars in a timescale of $\sim$1 Gyr when star formation proceeds in self-gravitating Giant Molecular Clouds (GMCs), while in a major merger this timescale is reduced by an order of magnitude \citep[e.g.][]{leroy08,solomon88}.  It is however reasonable to expect that intermediate systems exist, where the global (i.e. galaxy-wide averaged) star formation efficiency is found to vary within this framework \citep[as suggested also by e.g.][]{magnelli12}.  Such situations could occur either in the onset of a major merger as the transition between the two modes of star formation occurs, or in systems subjected to milder processes where part of the star formation is occurring normally in GMCs, while at the same time more violent star formation is taking place in other unvirialized regions.  For example, in the overlap region of the Antennae galaxies, \citet{wei12} observe a bimodal distribution of molecular clouds, finding both a population of GMCs typical of nearby galaxies and a population of unusually large, massive, and efficiently star-forming clouds.  \citet{boquien11} also previously reported significantly different star formation efficiencies in different regions of the interacting galaxy Arp 158, and \citet{papadopoulos12} argue in general that accurate modeling of the interstellar medium (ISM) of luminous infrared galaxies (LIRGs) requires consideration of both a central dense, efficiently star-forming phase and a more diffuse disc-like component where star formation proceeds at a more modest pace.  

Barred galaxies are examples of systems where the star formation process can be locally affected, as inward gas flows lead to increased central gas concentrations and star formation rates \citep[e.g.][]{sakamoto99,sheth05,masters12}.  Interactions with nearby companions can also induce gas flows that create high levels of star formation in the central regions of galaxies \citep[e.g.][]{mihos96,kewley06}.  Conversely, some galaxies appear to resist star formation, even in the presence of cold gas.  This includes early-type, bulge-dominated systems, where the presence of large reservoirs of atomic gas in the form of disks is not rare \citep[e.g.][and references therein]{grossi09,serra11}.  Some of these early-type galaxies even have significant, mostly compact, H$_2$ disks \citep[e.g.][]{wiklind95,knapp96,young11}, suggested to be less efficient at forming stars than similar gas disks in normal star-forming galaxies \citep{martig09,crocker12}. Indeed, several resolved studies point out that conditions in the ISM at the very centers of galaxies are different than in the disks, implying that the star formation efficiency is affected by the high stellar densities and/or the presence of AGN \citep{boker11,martin12,donovan12}.  

Here, we further investigate the link between structural parameters and star formation efficiency by looking directly at the cold gas component of galaxies.  Star formation is a local process, and is thus optimally studied with observations at the highest resolutions possible.  Our approach here is however different: by working with integrated galaxy properties and gas measurements, which can be measured in large unbiased galaxy samples, we can track the processes that have an influence on the global star formation efficiency of galaxies, and therefore the build-up of stellar mass at the present epoch.  These results can then be linked to high resolution studies of specific objects, to better understand the small scale physics responsible for the star formation efficiency variations.  We use data from the COLD GASS survey \citep[CO Legacy Database for the {\it Galex}-Arecibo-SDSS Survey;][]{COLDGASS1}, which was specifically designed to conduct such statistical analyses.  Compared with other samples which, while of comparable sizes, targeted specific populations \citep{young95,lisenfeld11,young11}, COLD GASS provides the full picture of the local galaxy population, over the mass range \mstar$>10^{10}$\msun.  

After describing the sample and the data in \S \ref{data}, we address in \S \ref{results} two questions left open in previous COLD GASS papers \citep{COLDGASS2,COLDGASS3}: (1) what is causing the molecular gas depletion time to vary across our sample? (2) what is the role of bulges in regulating and/or quenching star formation? As a sensor of the instantaneous star formation potential of galaxies, cold gas observations are a straightforward quantity that will help us answer these questions. A general interpretation of these results is presented in \S \ref{discussion}.    

All rest-frame and derived quantities in this work assume a \citet{chabrier03} IMF, and a cosmology with $H_0=70$\kms\ Mpc$^{-1}$, $\Omega_m=0.3$ and $\Omega_{\Lambda}=0.7$.

\section{Sample and Data}
\label{data}

We have been conducting a full suite of surveys to characterize the stellar and gaseous contents of massive galaxies, in a sample selected to be representative of the entire galaxy population in the local Universe with \mstar$>10^{10}$\msun.  The sample is selected in the redshift range $0.025<z<0.050$ from the SDSS DR6 spectroscopic survey, in the area of sky that overlaps with the GALEX MIS \citep{martin05} and the ALFALFA HI survey \citep{alfalfa1,haynes11}.  Full details regarding the sample selection are given in \citet{GASS1,GASS2}.  The GALEX and SDSS DR7 photometry was reprocessed using a matched-aperture technique, in order to derive accurate colors, star formation rates, and stellar masses \citep[see][]{GASS1,wang10,COLDGASS1,COLDGASS2}.  The star formation rates (SFRs) are obtained by fitting the photometry in the seven optical-UV bands to a set of model galaxies derived from the \citet{bc03} code with a range of ages, star formation histories, metallicities and dust attenuations.  For further details, see \citet{COLDGASS2}.  

We also use in the analysis two structural parameters often derived from SDSS data to quantify the morphology of the galaxies. The first is the stellar mass surface density, defined as:
\begin{equation}
\mu_{\ast}=\frac{M_{\ast}/2}{\pi R_{50,z}^2},
\end{equation}
where $R_{50,z}$ is the radius in kpc encompassing 50\% of the Petrosian flux in the SDSS $z-$band.  The second parameter, concentration index, is defined as $C\equiv R_{90,r}/R_{50,r}$.  It has been shown to correlate well with bulge-to-total ratios as determined by two-dimensional bulge/disk decompositions \citep{weinmann09}. 
 
Out of the parent population of galaxies matching these selection criteria, a subset was randomly selected for cold gas observations.  As detailed in \citet{GASS1}, atomic gas measurements are obtained with the Arecibo telescope.  The molecular gas masses come from observations of the CO(1-0) emission line with the IRAM 30-m telescope. The sample used in this study includes the 222 galaxies that were part of the first COLD GASS data release, and the additional galaxies observed until the completion of the survey in 2011 November, for a total of \ntot\ galaxies.  A full description of the instrumental setup, observing strategy and data reduction procedure is presented in \citet{COLDGASS1}.   We describe in \S\S \ref{apcorr} and \ref{gasmass} the aspects of the process that differ from those used in previous COLD GASS studies.

\subsection{Sample definitions}
\label{sample}

\begin{figure}
\epsscale{1.2}
\plotone{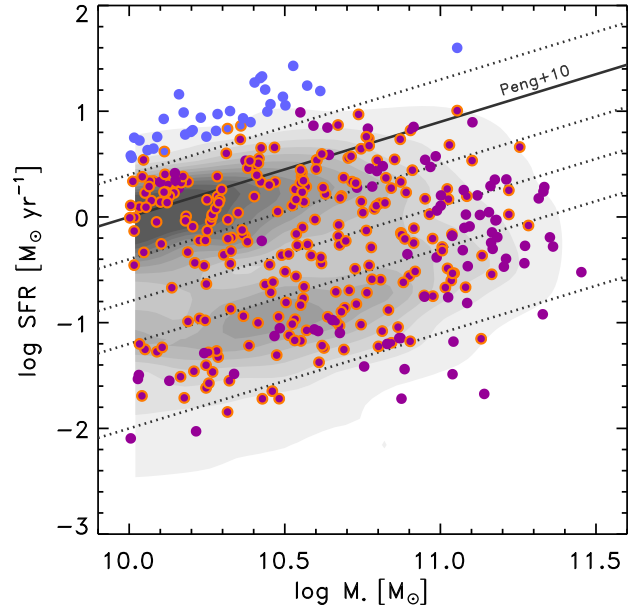}
\caption{Distribution of the sample in the SFR-\mstar\ plane.  The contours show the full unbiased SDSS parent sample or 12006 galaxies, and the filled colored circles are galaxies with CO(1-0) measurements.  The magenta symbols represent the galaxies in the main sample, the blue symbols the high SFR/\mstar\ galaxies.  We will refer to all galaxies with a CO measurement as the {\it full sample}.  We also define a {\it representative sample}, by picking a subset of the full sample that matches the distribution of the parent sample in this SFR-\mstar\ plane.  An example of a representative sample is shown by the open orange circles.  In this figure we also show the SFR-\mstar\ relation determined by \citet{peng10}, $\log{\rm (SFR)}=0.9(0.1\log M_{\ast}-1.0)$, and we define by the different lines six subsamples based on their distance from this star formation main sequence.    \label{MS}}
\end{figure}

\subsubsection{The full sample}

The COLD GASS sample is shown in Figure \ref{MS} in the SFR-\mstar\ plane.  Since the sample is purely mass-selected, it includes both star-forming and passive galaxies.  In order to quantify gas scaling relations across the entire galaxy population, the sample selected for gas observations (both HI and CO) is constructed to have a flat \mstar\ distribution \citep{GASS1,COLDGASS1}. Additionally, in order to fully explore the SFR-\mstar\ plane, we have targeted objects with high SFRs per unit mass.   In this study, we refer to the \ntot\ galaxies with CO measurements as the {\it full sample} (at the moment, HI data are available for 320 of these galaxies).   The full sample is meant to sample uniformly the entire SFR-\mstar\ plane, and as such has an excess of high mass galaxies, and of galaxies with high specific star-formation rates compared to a purely volume-limited sample.  

\subsubsection{The representative sample}

We also construct a {\it representative sample}, by selecting a subset of objects from the full sample that matches the distribution of the unbiased, volume-limited parent sample in the SFR-\mstar\ plane.  We adopted a grid in this 2-dimensional plane with cells of size 0.125 dex in \mstar\ and 0.2 dex in $\log{\rm (SFR)}$, and culled galaxies from the full sample so that the relative fraction of the sample in each grid cell matched that in the parent sample.  The distributions of other important quantities such as concentration index and stellar mass surface densities, are also statistically equivalent between the parent sample and this representative subsample (Kolmogorov-Smirnov probability, KS$_{prob}$, greater than 0.8 that the $C$ and \must\ distributions are drawn from the same parent sample).   We construct a set of 50 representative subsamples, an example of which is shown in Figure \ref{MS}.   In the remainder of this paper, we explicitly mention whether the full or the representative sample is used.  When the representative sample is used, we are able to assess the robustness of the results by comparing the significance of the results across the set of 50 repetitions. 

\subsubsection{The control samples}
In \S \ref{results}, we investigate different mechanisms that may affect star formation efficiency.  Because gas fractions and the depletion time depend on several global galaxy properties \citep{GASS1,GASS6,COLDGASS1,COLDGASS2}, it is not trivial to find a proper reference sample for any sub-population we wish to study (for example, galaxies with strong stellar bars or hosting AGN).   In these cases, we define a {\it control sample} of galaxies matched in some quantities, such that the effects of the mechanism under study on the depletion time can be assessed, everything else being equal.  Unless stated otherwise, we match the control samples on \nuvr\ color and stellar mass surface density, \must, using the technique described above.  The control sample therefore shares the same distribution as the sample of interest in the the 2D plane defined by \nuvr\ and \must, the two quantities upon which atomic and molecular gas contents depend the most \citep{GASS1,GASS6, COLDGASS1}.   For this reason, we must redefine a control sample each time we investigate a new sub-population of galaxies.

\subsection{Aperture corrections}
\label{apcorr}

To correct the CO fluxes measured in the 22\arcsec\ beam of the IRAM telescope to the total flux, we derive an aperture correction for each galaxy.  In the first COLD GASS data release, we used a small sample of nearby galaxies with resolved CO maps to derive a mean relation for the aperture correction as a function of $D_{25}$, the SDSS $g-$band optical diameter.  This method has the advantage of showing the range of aperture corrections that are possible at any given disk size $D_{25}$, given the real variation in radial molecular gas profiles within the reference sample.  However, it does not take into account the specific properties of each galaxy, including its disk inclination and light profile.    

In this paper, we therefore use a new prescription to correct fluxes, similar in spirit to the method used by \citet{lisenfeld11}.   For each galaxy with a CO measurement, we create a model galaxy having an exponential molecular gas distribution with a ``half-light" radius corresponding to $r_{50,SFR}$, the radius encompassing 50\% of the star formation as measured in the SDSS/{\it GALEX} photometry.  This choice is based on the observation that CO and SFR distributions trace each other well in nearby star-forming galaxies \citep{leroy09}, even in the HI-dominated outer disc regions \citep{schruba11}.  The model is then inclined as the real galaxy, and we simulate an observation with the 22\arcsec\ IRAM beam.  The ratio between this simulated observation and the total flux of the model galaxy is adopted as the aperture correction.  For the more face-on systems, these values are perfectly consistent with the previous estimates.  Only in the most inclined systems do we find aperture corrections on average 15\% smaller.  

The largest source of uncertainty in this process is the assumption that the gas profiles are exponential.  Resolved maps of CO in nearby galaxies indicate that this assumption is generally valid: molecular gas profiles are observed to be exponential, with the CO and SFR scale lengths equivalent, both of which being proportional to the stellar scale lengths \citep{young95,nishiyama01,leroy08}, albeit with some scatter especially at small radii \citep{regan01,bigiel12}. At fixed disk inclination, we observe differences of at most 15-20\% in the values of \mh\ obtained with aperture corrections derived from the method described above, and aperture corrections measured from the azimuthally-averaged observed SFR profile of each galaxy.  Lisenfeld et al. (2011) estimate a 30\% uncertainty on their aperture corrections obtained from a similar disk modeling technique (see their Figure 3).  Based on our tests and this independent estimate, we adopt a general uncertainty of 20\% on the aperture corrections.   Accounting also for the uncertainties in the measurement of the CO line fluxes, in the flux calibration at the telescope, and in the conversion factor \xco\ (see \S \ref{gasmass}), we estimate a global error of $\sim0.3$ dex on our values of \mh, with the error budget largely dominated by the systematic uncertainty on \xco\ \citep{COLDGASS1}.

\subsection{Molecular gas masses}
\label{gasmass}

The other difference with the procedure described in \citet{COLDGASS1} is the choice of the CO-to-H$_2$ conversion factor, \xco.   We have previously adopted a Galactic value of 3.2 \msun (K \kms\ pc$^{2}$)$^{-1}$ for all the galaxies in our sample, which did not take into account the contribution of Helium.  We now include this correction, and therefore use \xco$=4.35$ \msun (K \kms\ pc$^{2}$) for normal star-forming galaxies.  Among the full sample used in this study, there are however galaxies undergoing major mergers.  In such cases, it has been suggested that a value of 1.0 \msun (K \kms\ pc$^{2}$)$^{-1}$ is more appropriate \citep{solomon97}, and we follow this convention.  

To avoid biasing the results of this study, it is preferable not to tie the choice between the Galactic and ``merger" \xco\ to  a visual classification.  Instead, we wish to use a quantitative measure of the properties of the interstellar medium to make this choice.  While infrared luminosity is directly connected to merging activity in the most extreme cases \citep[ultraluminous infrared galaxies and sub-millimeter galaxies, e.g.][]{solomon97,veilleux02,engel10}, there is no one-to-one correlation between a high infrared luminosity ($L_{FIR}>10^{11}L_{\odot}$) and the need for a low value of \xco, nor between morphological signs of merging and enhanced star formation activity \citep[e.g.][]{bushouse88,alonso06,dasyra06,cox08}.  However, the high densities found in mergers may  lead to both elevated dust temperatures and lower values of \xco\ \citep{narayanan11}.  We therefore choose dust temperature, as parametrized by the $S_{60\mu m}/S_{100\mu m}$ IRAS flux ratio, to guide our choice between Galactic and ``merger" \xco.  Based on a sample of galaxies for which independent measures of \xco\ are available \citep{boselli02}, we choose a value of $S_{60\mu m}/S_{100\mu m}=0.5$ as the threshold to determine \xco\ (Gracia-Carpio et al., in preparation).    We therefore apply a conversion factor of  \xco$=1.0$ \msun (K \kms\ pc$^{2}$)$^{-1}$ to all galaxies with $\log L_{FIR}/L_{\odot}>11.0$ and $S_{60\mu m}/S_{100\mu m}>0.5$, and the Galactic value of \xco$=4.35$ \msun (K \kms\ pc$^{2}$)$^{-1}$ otherwise.   Of the \ntot\ galaxies in the sample, 11 match this criteria and use the lower value of \xco.  We note that within our survey volume, IRAS was sensitive to galaxies with $\log L_{FIR}/L_{\odot}>10.7$, therefore this convention to assign \xco\ does not bias against part of the sample, even though only 15\% of the COLD GASS galaxies have good quality flux measurements at both 60 and 100 $\mu$m.   In \S \ref{appendix2}, we show that key results of this study are robust against this specific choice of \xco.

\section{Results}
\label{results}

\subsection{Depletion time variations across the SFR-\mstar\ plane}
\label{mainsequence}

\begin{figure*}
\epsscale{1.1}
\plotone{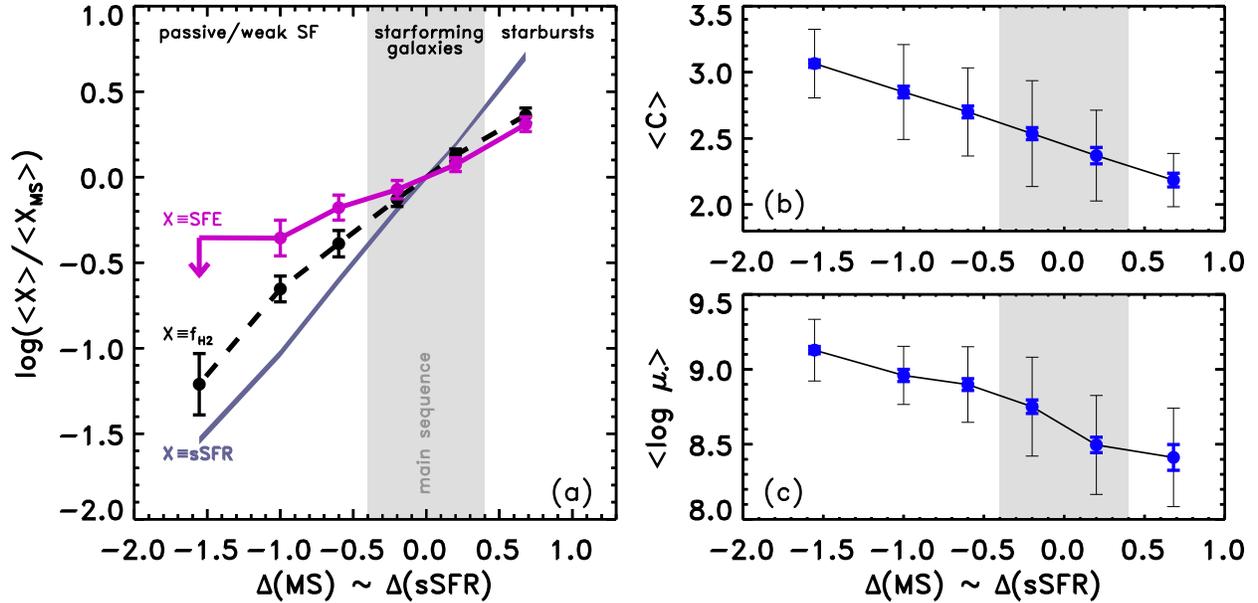}
\caption{{\it Left:} Mean depletion time (solid magenta line) and molecular gas fraction (dot-dashed black line) as a function of $\Delta$(MS), the offset from the sequence formed by star-forming galaxies in the SFR-\mstar\ plane (Fig. \ref{MS}).  The mean value of \tdep\ and \fgas\ in each bin is obtained by stacking all CO spectra, therefore including both detections and non-detections, and the error bars obtained from bootstrapping are 1$\sigma$ confidence intervals on the position of the mean.  The location of the main sequence is indicated by the gray area.  Both \fgas\ and the star formation efficiency (i.e. \tdep$^{-1}$) increase in galaxies with high sSFR and decrease in passive galaxies.  Comparing to the sSFR variation in the same bins (thick gray line), this figure shows that changes in sSFR across the SFR-\mstar\ plane are not caused only by variations in gas contents, but also by changes in the star formation efficiency.  {\it Right:} Mean concentration index ($<C>$, panel b) and stellar mass surface density ($<\log \mu_{\ast}>$, panel c) in the same bins of $\Delta$(MS).  The black error bars show the 1$\sigma$ dispersion in each bin and the thick blue error bars the uncertainty on the position of the mean as determined by bootstrapping.\label{MSoff}}
\end{figure*}

We have already reported variations in both molecular gas mass fraction (\fgas$ \equiv$ \mh $/$ \mstar) and depletion time (\tdep $\equiv$ \mh$/$SFR) across the COLD GASS sample in the mass range $10.0<\log M_{\ast}/M_{\odot}<11.5$ \citep{COLDGASS1,COLDGASS2}.    These observations are summarized in Figure \ref{MSoff}.   We however improve here on our previous results by using stacking to calculate mean values of \tdep\ and \fgas, allowing us to include both CO detections and non-detections in the analysis.  We compute these mean values in six bins of offset from the star-formation main sequence, calculated as:
\begin{equation}
\Delta{\rm (MS)}=\log{\rm (SFR)}-0.9(0.1\log M_{\ast}-1.0),
\end{equation}
adopting the main sequence definition of \citet{peng10}. The different $\Delta$(MS) bins are defined at fixed intervals, and illustrated in Figure \ref{MS}.  Since the \mstar-SFR relation is almost linear , $\Delta$(MS) is almost equivalent to variations in specific star formation rate (sSFR).  

To measure the mean depletion time of a given subsample, the stacking is done directly in ``depletion time units".   That is, all multiplicative factors required to convert an observed CO line flux into a \tdep\ value were applied to each spectrum prior to stacking.  This includes all the standard factors from the \citet{solomon97} prescription (their equation 2), but also the appropriate value of \xco\ (see \S \ref{gasmass}), SFR$^{-1}$, and a weight to correct for the flat stellar mass distribution of our sample \citep[the methodology for this is described in][]{COLDGASS1}.  After stacking these scaled spectra, the mean depletion time is obtained by integrating over the resulting line profile, selecting by hand the spectral window over which the flux is integrated to reflect the effective line width in the stacked spectrum.  This effective line width is seen to increase from $\sim300$ to $\sim500$ \kms\ going from low to high \must\ or $C$, as these quantities correlate with \mstar, which in turn correlates with circular velocity.   

Similarly, we obtain mean values of \fgas\ by stacking in ``mass fraction units" (i.e. using a \mstar$^{-1}$ scaling instead of SFR$^{-1}$).  This is the same technique used by \citet{fabello11a} to derive mean gas fractions from stacking, and further details can be found in that reference.  The only difference in methodology is that we do not weigh the spectra by their rms noise prior to stacking.  Applying such weights would cause severe biases as the rms noise in the individual IRAM spectra is correlated with the underlying properties of the galaxies as a result of our survey strategy.  Since we integrate until a galaxy is either detected in CO at $S/N>5$ or a sensitivity to a gas mass fraction \fgas=0.015 is reached, gas-rich galaxies and those at the smaller distances have spectra with higher rms than gas-poor and more distant objects.  Applying a weight derived form this rms noise would therefore heavily bias against gas-rich objects.

Figure \ref{MSoff}a shows that the mean value of both \tdep\ and \fgas\ varies as a function of $\Delta$(MS). Passive and weakly star-forming galaxies have low gas contents and long depletion times, while the most actively star-forming galaxies have higher gas contents and shorter depletion times.  This is consistent for example with the conclusion of \citet{combes94} that the increased star formation rate in interacting galaxies is due to both increased molecular gas mass and increased star formation efficiency.  The COLD GASS sample shows over a much broader range of galaxy types that changes in the molecular gas mass fraction cannot alone account for the full amplitude of sSFR variations; compare the black dashed line and the solid gray line in Figure \ref{MSoff}a.  The mean star formation efficiency must also vary, and the stacking analysis indeed reveals that it increases by a factor of $\sim6$ in the mean between passive galaxies in the red cloud and actively star-forming galaxies above the main sequence (see the magenta line in Fig. \ref{MSoff}a).  

What causes \tdep\ to vary between galaxies above, on and under the star formation main sequence? In \S \ref{appendix2} we argue that these changes are not caused by our choice of the CO-to-H$_2$ conversion factor.  Studies of large optical datasets both at low and high redshift however indicate that other important physical parameters vary smoothly as a function of distance from the star-formation sequence \citep[e.g.][]{schiminovich07,wuyts11}, and suggest that structural parameters quantifying bulge prominence are critical in setting the star formation properties of galaxies, or lack thereof.  We confirm in Figure \ref{MSoff}b,c that indeed our measures of ``bulginess", the concentration index $C$ and the stellar mass surface density \must, do vary with $\Delta$(MS), and in the next sections investigate the link between these structural parameters and star formation efficiency variations.

\subsection{Depletion time variations with galaxy structural parameters}
\label{sec_thresh}

\begin{figure*}
\epsscale{0.9}
\plotone{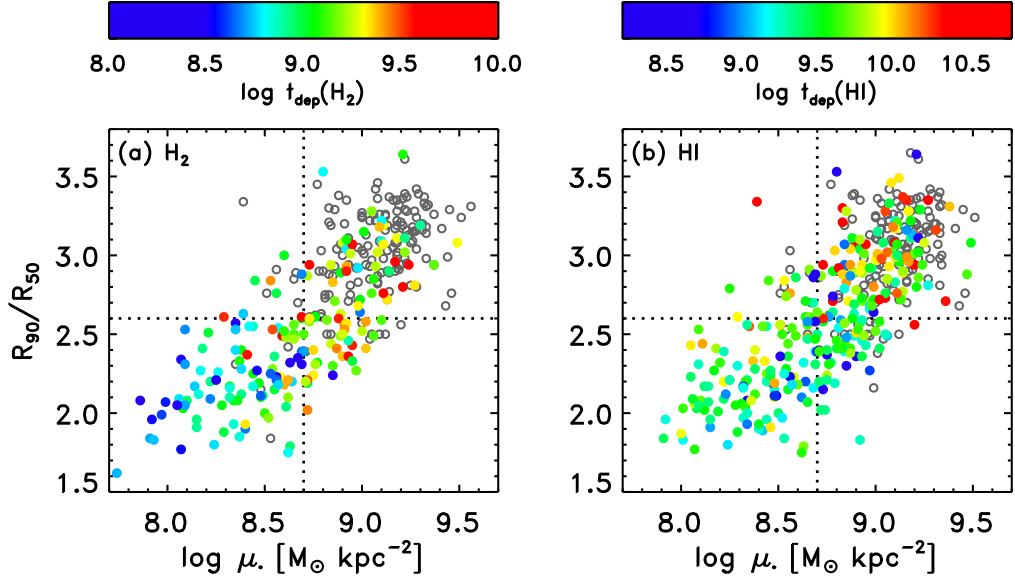}
\caption{Distribution of galaxies from the full sample in the \must-$C$ plane, color-coded by the depletion timescale of H$_2$ (panel a) and of HI (panel b).  In each case, non-detections are marked by open gray symbols, and the critical values of $C$ and \must\ above which most galaxies are non-detections are indicated by dotted lines.  Above these values (i.e. in the top-right quadrant of each plot), galaxies either are very gas-poor, or tend to have long cold gas depletion times.  \label{thresh}}
\end{figure*}

The importance of structural parameters (or morphology) in setting the gas properties of galaxies has previously been recognized.  For example, there are thresholds in $C$ and \must\ above which the detection rate of both HI and CO drops significantly \citep{GASS6,COLDGASS1}.  As also shown in \citet{COLDGASS3}, the $C-$\must\ plane is particularly efficient at separating the quenched population from the gas-rich, star-forming galaxies.  For this reason, we refer to the critical values of $C=2.6$ and $\mu_{\ast}=10^{8.7}$ \msun\ kpc$^{-2}$ as ``quenching thresholds".   These critical values, determined from the gas properties of the galaxies, agree well with similar thresholds in stellar population or morphological indicators.  For example, a concentration index of 2.6 corresponds to the transition between the late- and early-type galaxy populations, as reported by studies which compared this indicator to traditional visual morphologies, quantitative bulge/disk decompositions, and stellar population indicators \citep[e.g.][]{strateva01,shimasaku01,nakamura03,weinmann09,kauffmann03}.  Similarly, a stellar mass surface density of $3\times10^8$ \msun\ kpc$^{-2}$ was identified by \citet{kauffmann03} as marking the transition between galaxies dominated by young and old stellar populations.  In comparison, stellar mass is a poor quantity to separate the gas-rich from the gas-poor populations \citep{GASS1,COLDGASS1}.

In Figure \ref{thresh}, galaxies with cold gas mass fractions above the sensitivity limits of our surveys ($f_{gas}=1.5\%$) are color-coded according to their \tdep (panel a) or their \tdepHI (panel b).  This shows clearly that the galaxies with the longest values of \tdep\ and \tdepHI\ have high concentration indices and high stellar mass surface densities.  It therefore appears that bulge-dominated, high stellar mass density galaxies either (1) have very little cold gas, either atomic or molecular, or else (2) tend to be inefficient at converting into stars the molecular gas they do have. 

\begin{figure*}
\epsscale{0.9}
\plotone{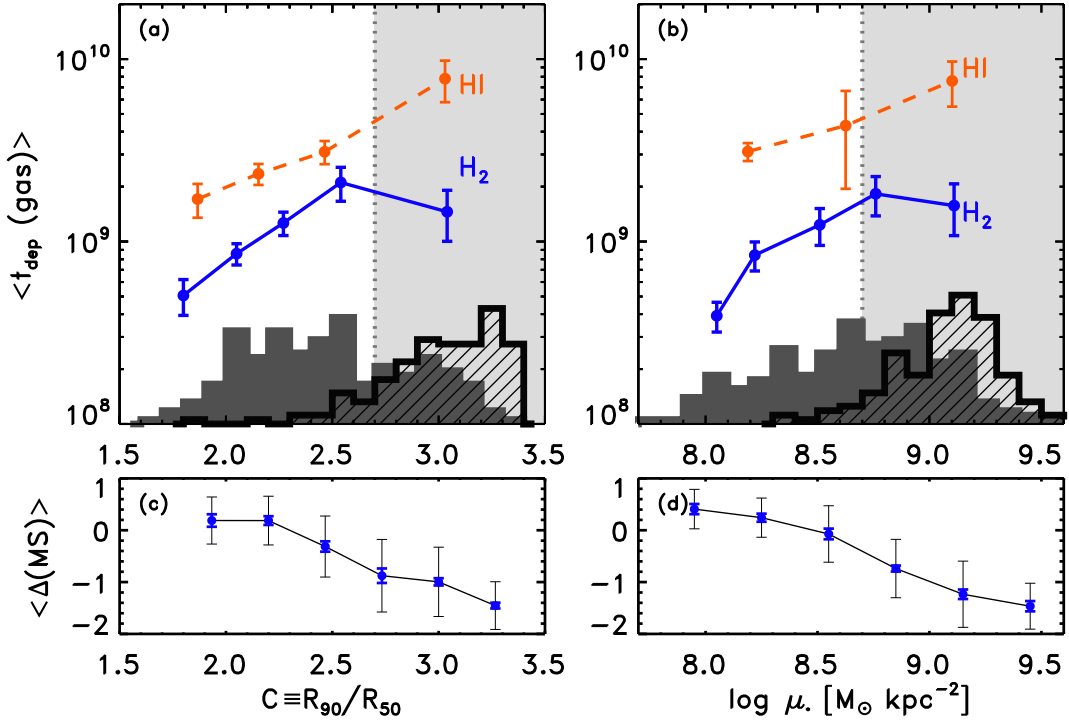}
\caption{Molecular and atomic gas depletion times as a function of concentration index (a) and stellar mass surface density (b) in the representative sample.   The mean values of \tdep\ and \tdepHI\ are obtained by stacking both detections and non-detections, and error bars are from bootstrapping (as in Fig. \ref{MSoff}).  The histograms at the bottom of each panel shows the distribution of CO detections (solid gray histogram) and non-detections (dashed black histogram), and the shaded regions represent the ranges of $C$ and \must\ values where galaxies are mostly bulge-dominated and gas-poor.  The detection rate as a function of \must\ and $C$ is similar for HI \citep{GASS1}.   In the bottom panels (c and d), the mean value of $\Delta$(MS) is shown as a function of $C$ and \must.  As in Figure \ref{MSoff}, the black error bars are the 1$\sigma$ dispersion in each bin and the thick blue error bars the uncertainty on the position of the mean as determined by bootstrapping.  \label{stacksCO}}
\end{figure*}

Care must be taken in interpreting this last observation.  Since at high \must\ and $C$ the detection rate of CO and HI is low, the galaxies we are detecting will tend to be the most gas-rich at fixed SFR, and therefore those having the longest depletion times.  While it can be unambiguously concluded from Figure \ref{thresh} that the longest values of \tdep\ and \tdepHI\ are only found in bulge-dominated galaxies, this ``Malmquist bias" means that we cannot conclude from Figure \ref{thresh} alone that the {\em mean} depletion time is longer in bulge-dominated galaxies.  To circumvent this issue, we can use stacking over representative samples of galaxies with CO and HI spectra in order to include both detections and non-detections in the analysis.  

The mean values of  \tdep\ and \tdepHI\  are shown as a function of $C$ and \must\ in Figure \ref{stacksCO}.  The stacking analysis confirms that both H$_2$ and HI depletion times are longer in galaxies with high $C$ and \must, even when including the CO and HI non-detections.  The specific behaviour of \tdep\ and \tdepHI\ as a function of $C$ and \must\ is however different.  The molecular gas depletion time increases steadily with $C$ and \must\ in the regime of disc-dominated galaxies located on or around the star formation main sequence ($|\Delta{\rm (MS)}|<0.4$, see Fig. \ref{stacksCO}c,d), and reaches a maximum value in the mean when reaching values of $C$ and \must\ characteristic of bulge-dominated galaxies.  In contrast, \tdepHI\ increases more slowly at $C<2.6$ and $\log \mu_{\ast}<8.7$, but then suddenly reaches a value close to a Hubble time for the bulge-dominated galaxies, located well below the star formation main sequence.   An interpretation of these trends is presented below in \S \ref{bulges}.

\subsection{Efficient star formation in dynamically active galaxies}
\label{highSFE}

\begin{figure}
\epsscale{1.2}
\plotone{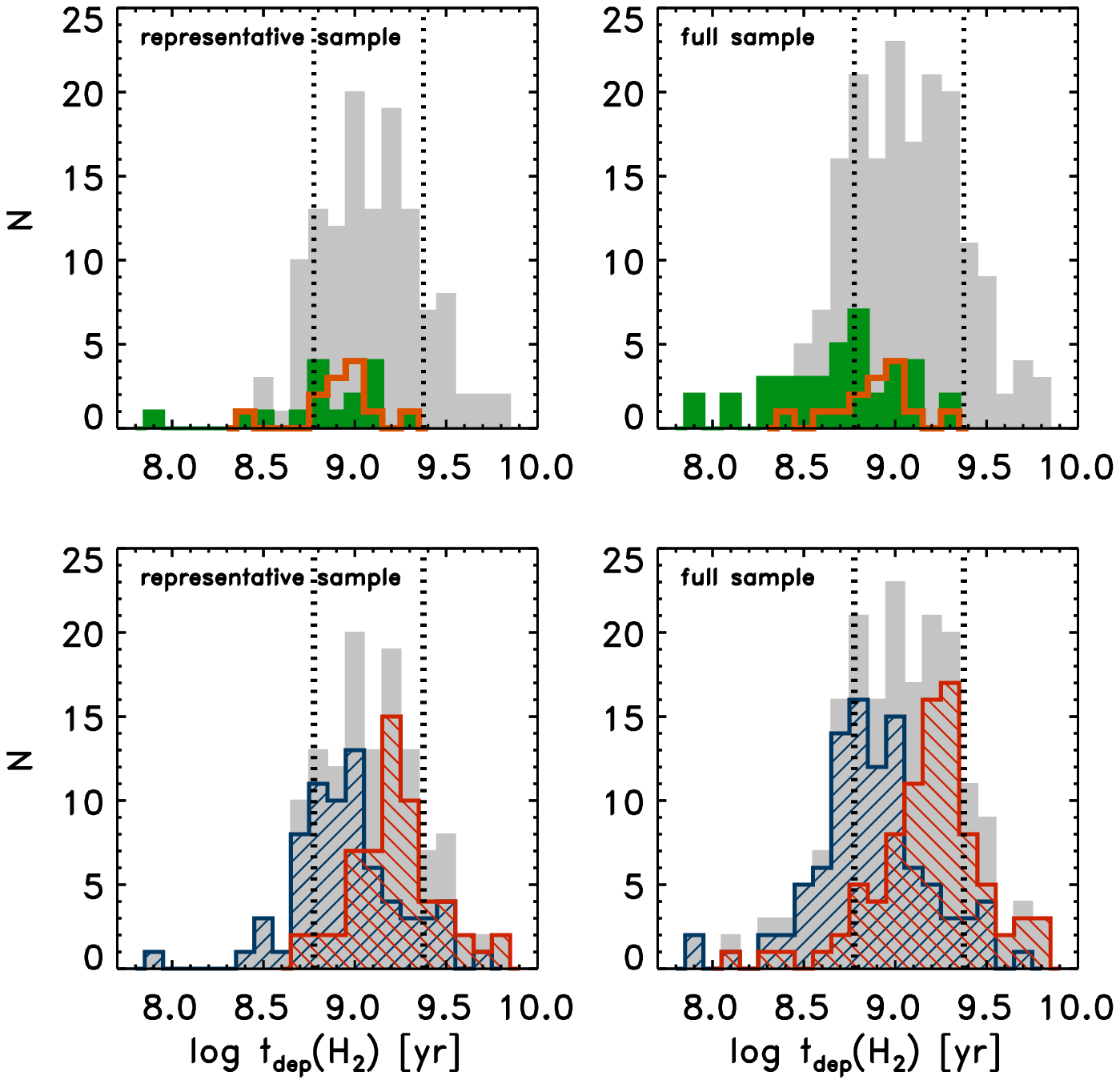}
\caption{Molecular gas depletion time distribution for all COLD GASS galaxies with $S/N_{CO}>5$ and \nuvr$<5.0$ (gray histograms).  {\it Top panels:} galaxies undergoing mergers or showing signs of strong morphological disturbances (i.e. high asymmetries) are represented by the green histogram, while galaxies with strong bars are shown as the orange histogram.  {\it Bottom panels:} depletion time distribution of the galaxies with \must$<5\times10^8$ \msun\ kpc$^{-2}$ (blue) and \must$>5\times10^8$ \msun\ kpc$^{-2}$ (red).     The left column shows results for the representative sample, while the right column displays the distributions in the full sample. \label{tdephists}}
\end{figure}

It is already established that in systems well above the main sequence of star-forming galaxies in the SFR-\mstar\ plane, \tdep\ is reduced by an order of magnitude compared to the normal star-forming population \citep[e.g.][]{gao04,bothwell10,genzel10,daddi10}.  In the local universe, these are the ultraluminous infrared galaxies (ULIRGs), most of which (if not all) are major mergers \citep{sanders96,veilleux02}.   For this reason, we speculated in \citet{COLDGASS2} that milder dynamical effects, such as minor mergers and bar instabilities, could lead to more modest, yet measurable variations in the global \tdep.    We have now also established that \tdep\ is longer in bulge-dominated galaxies than in star-forming disks, which may be interpreted in this same framework by an increased stability against star formation.  In this section, we address the question of the impact of interactions and bars on star formation efficiency, and come back to the role of bulges in \S \ref{bulges}. 

Whereas major mergers with mass ratios close to unity are very rare in the local Universe, the COLD GASS sample includes a fair number of objects undergoing milder dynamical perturbations which may be affecting their star formation efficiency.  The analysis here focuses on galaxies with bars, and on those displaying signs of interactions or morphological irregularities.    We use the classification of Wang et al. (2012) to identify galaxies with strong bars, defined as having ellipticity $e_{bar}>0.5$.  Not only is the identification of such bars robust, these are the kinds of bars that are linked to nuclear starbursts (see Wang et al. 2012 for the full details).  

Two different approaches are then used to identify interacting or otherwise disturbed galaxies.  First, we use a set of quantitative morphological parameters measured by \citet{GASS3} for the GASS sample, following the technique of \citet{lotz04}.  Secondly, we performed a visual inspection of all COLD GASS galaxies, and classified them as either being normal,  showing signs of interaction and/or morphological disturbances, or finally appearing to undergo a major merger.  All galaxies identified as major mergers are IRAS-detected, with significant infrared luminosities ($L_{IR}>10^{11} L_{\odot}$).   In what follows we use the visual classification, but in \S \ref{appendix1} we show that not only there is a significant overlap between the subsamples identified by the two methods, the results of the analysis are not affected by the classification technique.

\begin{deluxetable*}{lccccccccccc}
\tablecaption{Mean molecular gas depletion timescales\tablenotemark{a} in selected COLD GASS sub-samples \label{tdeptab}}
\tablehead{
\colhead{ } &  \multicolumn{3}{c}{Bulges}  &  \colhead{ } & \multicolumn{3}{c}{Bars}  &  \colhead{ } & \multicolumn{3}{c}{Interactions} \\
\cline{2-4} \cline{6-8} \cline{10-12}
\colhead{Sample} & \colhead{low \must}  & \colhead{high \must} &
\colhead{KS$_{prob}$} &  & \colhead{bars} & \colhead{control} & \colhead{KS$_{prob}$}  &  & \colhead{interact} & \colhead{control} & \colhead{KS$_{prob}$}}\\
\hline \\
\startdata
Full &  $0.96\pm0.44$ & $1.9\pm0.9$  & $<0.001$ & & $0.88\pm0.35$ &  $1.3\pm0.8$  &  0.26 & &  $0.63\pm0.41$ &  $1.1\pm0.6$  &  0.005 \\
Repr. &  $1.2\pm0.5$ & $2.0\pm0.9$  & $<0.001$ & & $0.87\pm0.24$ &  $1.2\pm0.6$  &  0.24 & & $0.95\pm0.44$ &  $1.1\pm0.5$  &  0.49 
\enddata
\tablenotetext{a}{The mean depletion times are in units of $10^9$ yr, and the errors quoted are the standard deviations around that mean value in any given subsample. }
\end{deluxetable*}

\subsubsection{Interacting galaxies}

In Figure \ref{tdephists}, the distribution of \tdep\ for the COLD GASS galaxies with reliable CO and SFR measurements is shown, for both the representative and full samples.   In the top panels, the populations of interacting/morphologically-disturbed and barred galaxies are highlighted.  In the bottom panel of Figure \ref{tdephists} we show that a simple stellar mass surface density cut at $\log \mu_{\ast}=8.7$ is also efficient in separating the short- and long-depletion time populations in both the full and representative samples.  The mean values of \tdep\ for the low and high \must\ samples and the statistical significance of the difference between the distributions are summarized in Table \ref{tdeptab}.  A KS test indicates that in both samples, low and high \must\ systems have \tdep\ distributions which have a probability $<0.001$ of being equivalent.  Since most of the morphologically disturbed galaxies have low stellar mass surface densities, to reliably assess the impact of interactions and bars on depletion time we define control samples of galaxies matched in \must\ and NUV$-r$ color but showing no signs of disturbance. 

As seen in Figure \ref{tdephists} and Table \ref{tdeptab}, morphologically disturbed galaxies in the {\it full sample} have significantly shorter depletion times than objects in the control sample (KS probability of 0.005 that their depletion times are drawn from the same distribution).  \citet{combes94} similarly observed the depletion time to be a factor of 2 shorter in galaxy pairs.  In \S \ref{discussion} we argue that the short \tdep\ values in interacting/disturbed galaxies come from a change in the fraction of the gas that is found in the densest phase, due to the pressure increase.  In the {\it representative sample} however, the depletion times of disturbed galaxies and their control are equivalent.  This is consistent with previous findings that mergers cannot all be identically associated with episodes of active star formation \citep[e.g.][]{dimatteo07}, either because some interactions may not drive gas inwards and compress it, or else because there is a delay between the occurrence of the signposts of morphological perturbations and starburst events.

\subsubsection{Barred galaxies}

Figure \ref{tdephists} also suggests that barred galaxies have shorter-than-average depletion times.  This is confirmed by the comparison with the control sample in both the {\it full} and {\it representative} samples, although the difference is small (a factor of 1.5), and the \tdep\ distributions only marginally different (KS$_{prob}$=0.25).  Several studies indicate that bars can drive gas inward and lead to locally enhanced nuclear star formation efficiency \citep[e.g.][]{sakamoto99,sheth05}.  In a detailed analysis of the barred population in the GASS parent sample,  Wang et al. (2012) find a clear link between bars and increased central star formation in galaxies with \mstar$>3\times10^{10}$\msun\ and \must$<3\times10^8$\msun\ kpc$^{-2}$, suggesting that bars are critical in the build-up of bulges, and possibly also in the quenching of star formation \citep[see also][]{masters11}.   Our observations of only a mild increase of the {\it global} star formation efficiency in barred galaxies however suggests that bar instabilities do not significantly affect the star formation budget in the local Universe.   Just as in the case of interacting galaxies, the week dependence of \tdep\ on the presence of strong stellar bars may be due to the fact that we average over systems at different stages of the gas inflow process.  Indeed, \citet{jogee05} show that the depletion time of the molecular gas in barred galaxies can be both long or short, depending whether a galaxy is observed in early stages of the bar-driven inflow, or at a later stage when most of the molecular gas is in the circumnuclear region.

\subsubsection{The global star formation law}

\begin{figure*}
\epsscale{0.9}
\plotone{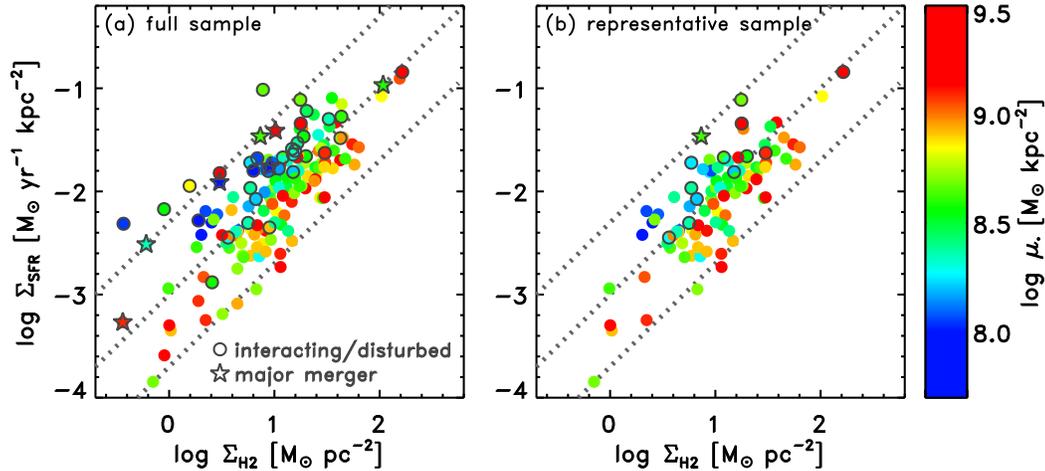}
\caption{Kennicutt-Schmidt star formation relation from COLD GASS.   Galaxies are color-coded by stellar mass surface density.  The open symbols outline the position of ``dynamically active" galaxies, and the dotted lines indicate constant depletion time values of 200 Myr, 1 Gyr and 5 Gyr, from top to bottom.   \label{KS}}
\end{figure*}
     
To assess whether the \tdep\ variations are caused by systematic changes in the global surface density of the molecular gas, the COLD GASS sample is shown in the Kennicutt-Schmidt (KS) plane in Figure \ref{KS}.  Surface densities are computed assuming that star formation and molecular gas are distributed similarly \citep{leroy08}, with a ``half-light" radius, $r_{50,SFR}$, measured for each galaxy as part of our standard photometry pipeline \citep{wang10}.   Results are shown for both the full and representative samples.  With the full sample one can assess the distribution of different galaxy types in the KS plane, while the representative sample allows the determination of an unbiased global star formation relation for local massive galaxies (\mstar$>10^{10}$\msun).  This relation is:
\begin{equation}
\log \Sigma_{SFR} = \Bigg \{ \begin{array}{l}
(0.86\pm0.21)(\log \Sigma_{H2})-(2.94\pm0.26) \\
(1.18\pm0.24)(\log \Sigma_{H2})-(3.30\pm0.28),  
\end{array} \label{KSeq}
\end{equation}
where the errors given for the slope and intercept are 3$\sigma$ formal fit errors, which approximates the total error including systematic effects and calibration uncertainties \citep{genzel10}.  As different authors use different fitting strategies, we present for comparison purposes the results of an ordinary least squares regression of $y$ on $x$ (top equation), and of a bisector fit (bottom equation, and our preferred method).  In both cases, the scatter of the residuals around the relation is $\sigma_{KS}=0.28$, and the slope is consistent with linear.  These values are in agreement with other derivations of the star formation law at the kpc-scale \citep{leroy08,blanc09,bigiel11,rahman12} and at high redshift \citep{genzel10,daddi10}. 

The \tdep\ variations observed across the COLD GASS sample do not come from a star formation law with significantly steeper slope than in other recent work, nor from systematic trends between $\Sigma_{H_2}$ and e.g. \must\ or interaction state.  Rather, the \tdep\ variations manifest themselves as structure within the scatter around a $\sim$linear KS relation, where specific populations, for example galaxies with high/low \must\, are preferentially located below/above the mean KS relation (Fig, \ref{KS}).  As expected from previous studies, the galaxies undergoing major mergers have short depletion times and mark the upper envelope of the distribution in the KS plane (Fig. \ref{KS}).  Mildly interacting/disturbed galaxies, having low stellar mass surface densities, then fall systematically above the mean KS relation.  

Additionally, galaxies with the highest stellar mass surface densities and concentration indices (i.e. the bulge-dominated galaxies) are found systematically below the mean KS relation, with long global molecular gas depletion times ranging between 1 and 5 Gyr.  A similar observation was made by \citet{crocker12} for a small sample of nearby early-type galaxies with detailed observations of multiple molecular gas lines.  In Figure \ref{KS}a, there are a few galaxies deviating from the rest of the population, having high stellar mass surface densities yet short depletion times.  These are however all merging systems, which may be in a transition phase and will ultimately lie below the mean KS relation after the starburst phase, as suggested by simulations \citep{bournaud11a}.   We come back to this discussion in \S \ref{bulges}.  

The main conclusions to be drawn from Figure \ref{KS} are that (1) in the {\it representative sample}, the ``merger branch" of the KS relation mostly disappears, confirming that very efficiently star-forming major mergers are rare events that do not contribute significantly to the star formation budget in the local Universe, (2) the global surface density of H$_2$ (estimated form our CO(1-0) line fluxes) is not the only parameter driving the star formation efficiency of galaxies, as we observe structure within the scatter about the mean KS relation, and  (3) the bulge-dominated galaxies, which have the longest molecular gas depletion times, are inefficient star-formers despite have $\Sigma_{H_2}$ values as high as spiral galaxies \citep[see also][]{young08,shapiro10}.


\subsection{The role of AGN}
\label{AGN}

As explained in \S \ref{intro}, most studies suggest that quenching is not strongly linked with stellar mass, but rather with a measure of ``bulginess".   Given the well-established correlation between supermassive black holes and bulges \citep[e.g.][]{haring04,magorrian98,merloni10,beifiori12},  AGN feedback as an important quenching mechanism is appealing \citep[e.g.][]{somerville08}, as it would explain why few galaxies without central mass concentration are quenched, while a very large fraction of the bulge-dominated systems are \citep[e.g.][]{bell08}.  

In this light, there has been work done to investigate the link between the presence of AGN and the gas content of galaxies, which should be most directly affected by the feedback.  However, \citet{ho08} and \citet{fabello11b} report no difference in the atomic gas properties of active and inactive galaxies, in large representative nearby samples.  This would seem to argue against the AGN feedback scenario presented above, although both studies point out that molecular gas, which is generally significantly more centrally-concentrated than HI, may be a more sensitive tracer.   Here, we use our IRAM CO observations to investigate whether the presence of an active nucleus influences the molecular gas contents of galaxies, in order to address the question of why bulge-dominated galaxies have on average longer \tdep.  As detailed below, we find that while AGN may affect the amount of molecular gas in their host galaxies, the star formation rate is also reduced, leaving the depletion time unchanged (or if anything {\it reduced}) as compared with a properly-matched control sample. 

We identify as active the COLD GASS galaxies located above the \citet{kauffmann03AGN} line in the BPT diagram \citep[the line ratio diagnostic plot first presented by][]{baldwin81}.  Of these, 6\% are Seyferts, the rest being either LINERs or composite systems where part of the line emission is from star formation.  In the reference sample of ``non-AGN" we include the star forming systems identified from the BPT diagram and the inactive galaxies which have no line emission ($S/N<3$ in H$\alpha$, [NII], H$\beta$ and [OIII]).  To compare fairly their molecular gas properties, we match subsamples of AGN hosts and reference galaxies, requiring that the two subsamples have equivalent distribution in \must\ and \nuvr\ color.  The results presented below are unchanged if instead we control for star formation activity in the central region (i.e. using the fiber D$_n(4000)$ instead of the global \nuvr\ color), or for another structural parameters such as \mstar\ or $C$.  

\begin{figure*}
\epsscale{0.9}
\plotone{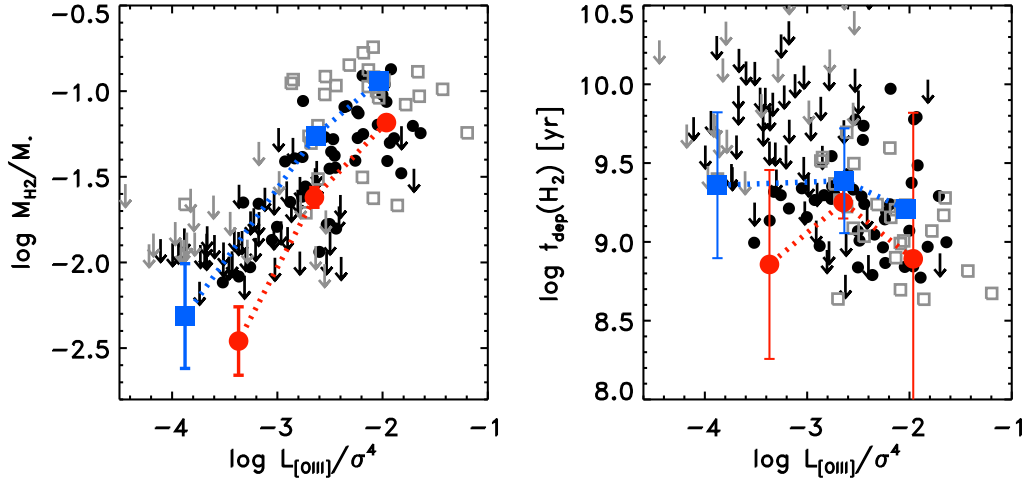}
\caption{H$_2$ mass fraction and depletion time as a function of the Eddington ratio, parametrized as $L$[OIII]$/\sigma^4$.  Individual CO detections and non-detections are shown, as well as mean values in bins of $L$[OIII]$/\sigma^4$ obtained by stacking (colored symbols, red: AGN, blue: control sample).  Galaxies in the AGN subsample (filled circles) have lower molecular gas mass fractions than galaxies in the control sample (squares), but their depletion times are equivalent if not shorter.  \label{Ledd}}
\end{figure*}

As a proxy for the Eddington ratio, we use the ratio $L$[OIII]$/\sigma^4$.  The luminosity in the [OIII] line, which we correct for extinction following \citet{wild07}, can be used as a proxy for the black hole accretion rate \citep{kauffmann03AGN}, while the black hole mass is proportional to $\sigma^4$ \citep{tremaine02}.  We use for $\sigma$ the stellar velocity dispersion measured in the SDSS fiber, corrected for aperture effects as detailed in \citet{graves09}.   Each subsample (AGN and control) is divided in three bins of $L$[OIII]$/\sigma^4$ and galaxies are stacked in order to obtain the mean values of \fgas\ and \tdep.  The results are shown in Figure \ref{Ledd}.  At all values of the Eddington ratio, the mean molecular gas mass fraction is lower in the AGN subsample, but the mean depletion times are consistent or even shorter than in the control.  

\citet{maiolino97} conducted a similar study, comparing the molecular gas properties of Seyfert and field galaxies.  They report no strong difference in the global H$_2$ contents, but an increased star formation efficiency in Seyfert 2 hosts.  Our very tentative observation of a shorter \tdep\ in AGN hosts may be the extension of what these authors are finding in the more powerful Seyferts, although a direct test of this is not possible due to the lack of overlap between the types of AGN probed by our studies and the different methodology in assigning control samples.  

The most important conclusion of this section in the context of this work is that the high incidence of AGN in bulge-dominated galaxies cannot explain why \tdep\ is longer at high concentration index and \must.  There does however appear to be a link between AGN and \fgas, but we caution that this observation pertains only to the type of AGN which are in our sample, typically LINERs accreting at a rate of at most $\sim0.01 L_{Edd}$.


\section{Discussion}
\label{discussion}

\subsection{Summary of observational results}

The key observations presented in this work can be summarized as follows:
\begin{enumerate}
\item Both the molecular gas mass fraction (\fgas) and depletion time (\tdep) vary across the SFR-\mstar\ plane.  Galaxies above the star-forming sequence have both higher \fgas\ and shorter \tdep.  Conversely, explaining the low specific star formation rates of galaxies below the star-forming sequence requires both \fgas\ to decrease and \tdep\ to increase. 

\item The bulge-dominated galaxies in the sample, identified by their large concentration indices or their high stellar mass surface densities ($C>2.6$, \must$>10^{8.7}$\msun\ kpc$^{-2}$) either have no cold gas down to the limit of our survey, or else are on average inefficient at converting into stars the molecular gas they do have.  Despite the fact that molecular gas in early-type galaxies has high surface densities just like in normal star-forming discs, we find their mean \tdep\ to be twice as long.  The atomic gas depletion time is also longer in bulge-dominated galaxies. 

\item At fixed \must\ and color, galaxies hosting AGN have lower molecular gas mass fractions than their control sample, at all values of the Eddington ratio probed by our sample.  There is tentative evidence that the depletion timescale of the molecular gas is either unchanged or lowered in the presence of an AGN, but not increased, indicating that active nuclei are not the main agent responsible for the long depletion times in bulge-dominated galaxies. 

\item Among the gas-rich, disc-dominated population, galaxies undergoing merging events or having disturbed morphologies have the shortest depletion times, on average a factor of 2 shorter compared to matched control samples.  Not all mergers or interactions are however associated with episodes of particularly efficient star formation.  

\item Global molecular gas depletion times are only mildly shorter in galaxies hosting strong stellar bars, everything else being equal.  Even though resolved gas studies show that bars can lead to enhanced nuclear star formation efficiency, the effect is not strong enough to significantly affect the {\it global} depletion time.  

\item We derive the star formation law for an unbiased sample of galaxies, representative of the local population with \mstar$>10^{10}$\msun.  This relation has a slope of $1.18\pm0.24$, consistent with previous studies; the \tdep\ variations manifest themselves as structure within the scatter around this best-fit relation, with bulge-dominated galaxies falling systematically below the mean relation.

\end{enumerate}

We now discuss these results, addressing the two main questions posed in the Introduction, and comparing them to previously published work.

\subsection{What is causing the global star formation efficiency to vary across the local massive galaxy population?}

Our results confirm previous observations that the transition from star forming to quiescent galaxies is correlated with the presence of a bulge, as quantified either by stellar mass surface density, Sersic index, velocity dispersion, or concentration index \citep[see also][]{GASS1,COLDGASS1,COLDGASS3}.   While there is still some debate as to which of these measures of  ``bulginess" correlates best with quiescence \citep[for different views on this, see e.g.][]{franx08,wuyts11,bell12,wake12}, various studies generally agree that stellar mass is a poorer diagnostic to separate star forming from quiescent galaxies.  Our results bring additional weight to these previous studies, since they are based on cold gas measurements, a direct tracer of the instantaneous potential of a galaxy to form stars, rather than on indicators such as color, which are averages over the different star formation history of each galaxy.   

Structural parameters such as \must\ and $C$ do not only correlate with quiescence, but also with star formation efficiency.  These physical quantities are therefore at the core of this question.  Only in galaxies with lower \must\ do we see the dynamical processes that can significantly increase star formation efficiency.  At the other extreme, high central concentrations may lead to reduced star formation efficiency or complete quenching of the star formation.   

The single quantity that may link these observations is the dense molecular gas fraction.  We speculate that while \tdep\ (measured from CO(1-0) line fluxes) varies across our sample, were we to measure a depletion time for the dense gas, it would show significantly less scatter.  A commonly used dense gas tracer is the HCN(1-0) line, with observations showing that the HCN(1-0)/CO(1-0) ratio is increased in efficiently star-forming galaxies like LIRGs \citep{gao04,juneau09,gracia08,garcia12}, and possibly reduced in early-type galaxies \citep{crocker12}.  We caution however that although the HCN/CO ratio appears to be a signpost for the star forming dense gas fraction, it is not a perfect tracer \citep[e.g.][]{gracia06}. Indeed, HCN is a minor species subject to the abundance of Nitrogen and the details of the chemical network \citep[e.g.][]{sternberg94,usero04,meijerink07}, and its (1-0) transition is sensitive to the specific physical excitation conditions with mid-infrared pumping possibly playing a role \citep{aalto95}. Other possibilities to trace the dense star-forming gas would be the HCO$^{+}$(1-0) transition, or higher rotational transitions of e.g. CO, HCN or HCO$^{+}$. 

Theoretical and numerical work also supports this picture. \citet{krumholz07} find that a line with a high critical density like HCN(1-0) traces well the efficiently star forming gas between galaxies with different global properties.  On the other hand, they find that a lower density tracer like CO(1-0) tracks regions with various star formation properties from galaxy to galaxy.  With high resolution hydrodynamic simulations, \citet{bournaud11a} find the probability distribution function (PDF) of the local gas density in mergers to be broader and shifted to higher densities (or perhaps to have a different functional form), compared to the mostly log-normal PDFs found in normal star-forming galaxies \citep{wada01}.  They conclude that these variations in the dense gas fractions at different stages of a merger cause galaxies to go from having ``normal" \tdep\ before the interaction begins, to having short \tdep\ at the peak of the merging activity (resulting in the merger sequence in the Kennicutt-Schmidt plane).  The end product of these mergers, early-type systems with some amount of dense molecular gas left, are somewhat inefficient at forming stars and have longer than average \tdep.  This scenario explains well the location of the various galaxy types on the COLD GASS KS relation (Fig. \ref{KS}).

\subsection{What is the role of bulges in regulating/quenching star formation?}
\label{bulges}

In early simulation work, \citet{ostriker73} pointed out that galaxies with significant spheroidal components are more stable against fragmentation than pure disks.  Since fragmentation introduced by gravitational instabilities is a key element of star formation,  the natural conclusion may be that bulges can therefore reduce star formation efficiency.  This can be understood in terms of the stability parameter, $Q$ \citep{toomre64}, which in the case of gas$+$stars systems can be extended to include the two components \citep[e.g.][]{jog84,elmegreen95,romeo11}, as confirmed by observations \citep{wong02,leroy08,hitschfeld09}. Specifically, a gas disk, which would be otherwise unstable, can be made stable by a significant stellar bulge.   Under the framework presented above, long depletion times in bulge-dominated galaxies would therefore result from a gas density PDF shifted to lower densities than in normal star-forming galaxies.   We note that the effect of a stellar component on the stability of a gas disk could however be reversed in galaxies with lower stellar masses than those in the COLD GASS sample, when stars are distributed in a disk without a significant central concentration.  In that case, higher stellar mass surface densities could translate into a shorter depletion time, as argued for example by \citet{shi11}. 

As seen in Figure \ref{KS}, bulge-dominated galaxies fall systematically below the mean molecular Kennicutt-Schmidt relation, indicating that at fixed gas surface density, they are less efficient at forming stars.  The observation is in line with the scenario presented above, in which massive stellar bulges can stabilize gas disks, therefore reducing the star formation efficiency.  Although this mechanism is not sufficient in itself to make star-forming galaxies ``red and dead", it may be efficient at keeping already-quenched galaxies on the red sequence, by preventing further star formation even in the event of late-time cold gas accretion  \citep{martig09}.    In massive galaxies, radio-mode AGN feedback may be the most important agent in keeping galaxies on the red sequence \citep[e.g.][]{gabor11,vandevoort11}, but the contribution of the bulge in the reduction of the star formation efficiency could however be an important effect in lower mass galaxies where AGN are not prevalent. 

Interestingly, we also find the longest values of \tdepHI\ in galaxies above the ``quenching thresholds" ($C>2.6$ and \must$>10^{8.7}$\msun\ kpc$^{-2}$, see Fig. \ref{thresh}), although in this case the reason may be different, and the different behavior of \tdep\ and \tdepHI\ as a function of \must\ hints at changes in the bottleneck of the star formation process across the local galaxy population.  In high density bulge-dominated galaxies, the atomic gas depletion time increases significantly.  This suggests that these galaxies have large reservoirs of HI gas, presumably located in the outer disk, that do not participate in the star formation process.  In this regime, the bottleneck for star formation is the transport of this atomic gas to regions of higher density where the atomic-to-molecular conversion (and hence star formation) takes place efficiently.   At stellar mass surface densities characteristic of disc-dominated galaxies, \tdepHI\ has a mean value of $\sim3$ Gyr with large galaxy-to-galaxy variations \citep{GASS2}, but increases slowly with \must\ and $C$ similarly to \tdep.  On average, there is therefore no accumulation of atomic gas, the conversion to CO(1-0)-detectable molecular gas proceeding unimpeded.  The \tdep\ variations then come about as galaxies have different fractions of their molecular gas in efficiently star-forming dense cores.  This interpretation is supported by our observation that even among galaxies with low stellar mass surface densities, the ones undergoing merging, or to some extent those hosting a strong stellar bar, have the shortest CO-based \tdep. 

Since bulge-dominated galaxies preferentially host AGN, it has been suggested that the correlation between quenching and the presence of a bulge may be a consequence of the feedback from the AGN.  If this were the case, there should be differences in the gas contents of galaxies with and without AGN, everything else being equal.  Compared to a control sample matched in \must\ and color (either global \nuvr\ or central D$_n$(4000)), galaxies hosting an AGN with $-4<\log L_{\rm [OIII]}/\sigma^4<-2$ have lower molecular gas mass fractions than inactive galaxies.  This observation indicates a possible connection between AGN activity and the quenching of star formation, although our sample does not include the stronger AGN which are likely to contribute most to this process.  Galaxies with and without AGN in our matched subsamples do not have statistically different mean values of \tdep, however, suggesting that the AGN with modest accretion rates are not responsible for the long depletion times in bulge-dominated galaxies.  There is very marginal evidence that \tdep\ is even reduced in AGN hosts, although the difference is not statistically significant in our sample which includes mostly LINERs.  The non-result is in agreement with \citet{juneau09} that report no difference in the HCN/CO ratio between galaxies with and without AGN, which in the framework of this discussion could explain why we do not observe variations in the CO-based values of \tdep\ for AGN hosting galaxies.  The disconnect between nuclear properties and outer disk star formation explains why there is no correlation between AGN activity and the generally extended HI contents of the host galaxies \citep{ho08,fabello11b,diamond12}.

\subsection{Comparison with results at the kiloparsec scale}

One of the main results of this work is that global structural parameters, in this case concentration index and stellar mass surface density, are critical in establishing both the bimodality between gas-rich star-forming and gas-poor quiescent galaxies, and in setting the efficiency of star formation.  Star formation however is not purely a global process, as it depends strongly on small-scale physical conditions, which our integrated measurements cannot track.  Significant work has however gone into measuring gas and star formation in nearby galaxies at the (sub-)kiloparsec scale, with results, if taken at face value, that may appear at odds with the conclusions of this study \citep[e.g.][who present evidence for a constant depletion time in nearby galaxies]{bigiel11}. 

The relation we observe between \tdep\ and the {\it global} stellar mass surface density may not apply at small scales, although Figure 15 of \citet{leroy08} indicates that the {\it local} mean depletion time in kpc-sized regions of nearby spirals increases by a factor of two over one decade in \must\  \citep[a similar effect can be seen in Figure 2 of][]{rahman12}.  The global values of \must\ may not track the local conditions in individual sites of star formation, but rather the probability of incidence of various dynamical processes which can influence whether star formation proceeds ``normally" in virialized GMCs or rather in dense super star-forming complexes.   

In a study of the resolved star formation law, \citet{schruba11} argue that star formation efficiency does not vary significantly within individual galaxies, with however significant galaxy-to-galaxy variations.  Even though this is consistent with the observations presented here, the interpretation is fundamentally different.  Indeed, \citet{schruba11} ascribe the \tdep\ variations to a metallicity dependence of the conversion factor \xco, while our analysis suggests that they are caused by actual changes in star formation efficiency.   As detailed in the appendix, we do not find any evidence for metallicity or extinction to be the main driver of the observed \tdep\ variations.  In the stellar mass range of the COLD GASS sample ($>10^{10}$\msun), the mass-metallicity relation is flat \citep{tremonti04}, implying that quantities such as \must\ and $C$ do not correlate with metallicity.  However, toward the low mass end of our sample, and even more so at \mstar$<10^{10}$\msun, the M$_{H_2}/$SFR ratio may very well be modulated simultaneously by variations in \xco\ and actual changes in star formation efficiency.  Even after accounting for metallicity effects on \xco, high star formation efficiencies have indeed been reported for several low mass, low \must\ local galaxies where high resolution measurements of both gas and star formation are possible, for example M33 \citep{gardan07,gratier10b}, IC10 \citep{leroy06}, and NGC6822 \citep{gratier10b} \citep[see however the counter argument in the case of the SMC in][]{bolatto11}.  Disentangling the two effects will require both large statistical samples of low mass galaxies akin to COLD GASS, and resolved gas maps of galaxies spanning a broader range of properties.  

The main difference between our global results and those of studies focussing on the (sub-)kpc scale may rather be more methodology-based.  As pointed out correctly by \citet{rahman12}, resolved studies measure \tdep\ in regions of the galaxies where CO is detected, but studies like ours include the total gas mass and the total star formation rate.    The resolved measurements therefore teach us about star formation in the gas-rich GMCs and the global measurements about the total timescale for star formation, including any diffuse gas or star formation component.  In particular, the global values of \tdep\ include star formation coming from the outer disks, regions where CO is in general not detected in resolved maps and therefore not included in resolved studies.  Since star formation is known to be very inefficient in these outer regions \citep[\tdepHI$>t_{Hubble}$,][]{bigiel10}, the relative contributions of outer disk star formation to the total value may be responsible for some of the galaxy-to-galaxy star formation efficiency variations, and the difference in absolute value of the mean depletion time reported in different studies (in general a factor of $\sim2$ higher in resolved studies).  For example, \citet{bolatto11} find \tdep$\sim1.6$ Gyr at the kpc scale in the SMC, but \tdep$=0.6$ Gyr for the entire galaxy, in line with the values we find in galaxies with low stellar mass surface densities. There are however arguments against the ``outer disk effect" interpretation.  First, there is evidence that the molecular star formation law extends even in the HI-dominated outer regions of normal star-forming discs \citep[e.g.][]{heyer04,schruba11}.  Secondly, the metallicity gradients of galaxies in the stellar mass range \mstar$>10^{10}$\msun\ are remarkably flat  out to the optical diameter \citep{moran12}, such that there is no evidence that the CO measurements of COLD GASS galaxies are subject to radially-dependent \xco\ variations. 

High resolution mapping of CO in a broad range of galaxies will be necessary to further make the connection between local and global measurements.  As the latter are most commonly available for high redshift systems, understanding this relation is critical.  Surveys like COLD GASS, offering homogeneously measured quantities over large, representative samples of galaxies are essential to identify the key populations where star formation efficiency is modulated, allowing us to understand the physical processes that are responsible for regulating the star formation budget of the local universe. For example, while bars can lead to nuclear starbursts and the formation of bulges, they are not responsible for increasing significantly the star formation efficiency over the local galaxy population.  Galaxy interactions and mergers however do have that effect.  Now that statistically-large samples are starting to be collected at high redshifts \citep[][and in prep.]{tacconi10}, it will also be possible to track over cosmic time the relative contributions of different mechanisms such as bar instabilities, bulge stabilization, interactions/mergers and AGN feedback in the regulation of star formation.

\acknowledgments
This work is based on observations carried out with the IRAM 30 m telescope. IRAM is supported by INSU/CNRS (France), MPG (Germany), and IGN (Spain).    We thank Fr\'ed\'eric Bournaud and Benjamin Magnelli for useful comments.

\bibliographystyle{apj}


\appendix

\section{A. Morphological classifications}
 \label{appendix1}

One of the objectives of this work is to evaluate the impact of interactions on the star formation efficiency, requiring us to first identify the subset of COLD GASS galaxies undergoing such events. We do not wish to restrict the analysis to systems that can be classified as major mergers or that have massive companions, of the type that could be identified in the SDSS spectroscopic database.  Instead, we choose to identify all the galaxies showing signs of morphological disturbances.  

\citet{wang10} measured for all the galaxies in the GASS/COLD GASS survey two different quantitative morphology parameters, based on the approach of \citet{lotz04}.  The asymmetry parameter ({\it Asym}) provides a measure of the difference between the SDSS $g-$band image and a copy of the same image rotated by 180$^{\circ}$.  The larger the value of {\it Asym}, the more asymmetric the galaxy.  The second is the clumpiness parameter ({\it Clump}), computed from the difference between the original $g-$band image and a copy of the same image smoothed to a scale of 20\% of the Petrosian radius.  Galaxies with clumpier morphologies therefore have larger values of {\it Clump}.  Two additional parameters are commonly used to quantify the morphologies of galaxies: the Gini ($G$) and $M_{20}$ parameters. The Gini coefficient resembles the concentration index, but does not require the use of aperture photometry or the identification of the galaxy center, since it instead relies on the distribution of pixel values \citep{abraham03}.  It therefore performs well in very clumpy systems with ill-defined centers of light.  The $M_{20}$ index is also akin to concentration, but again considers the full pixel distribution and therefore does not rely on the use of regular apertures \citep{lotz04}.   The SDSS $g-$band images are used to measure $G$ and $M_{20}$ for all COLD GASS galaxies. 

To assess the performance of these four indicators at isolating the morphologically-disturbed systems we wish to study, we also perform a visual classification.  Galaxies are marked as ``disturbed" if they satisfy one of three criteria: (1) they have a companion within 1$^{\prime}$ (i.e. 30-60 kpc in our redshift interval of $0.025<z<0.050$), (2) tidal features are visible, or (3) the stellar disks are lopsided or show strong asymmetrical features.   The classification was done ``blindly", without any knowledge of the value of the morphological parameters or any other physical properties of the galaxies that were inspected.  AS performed the classification twice, randomizing the order of the galaxies each time to avoid ``fatigue biases".  Only galaxies consistently classified as disturbed in the two runs are flagged as such.  

\begin{figure}
\epsscale{0.7}
\plotone{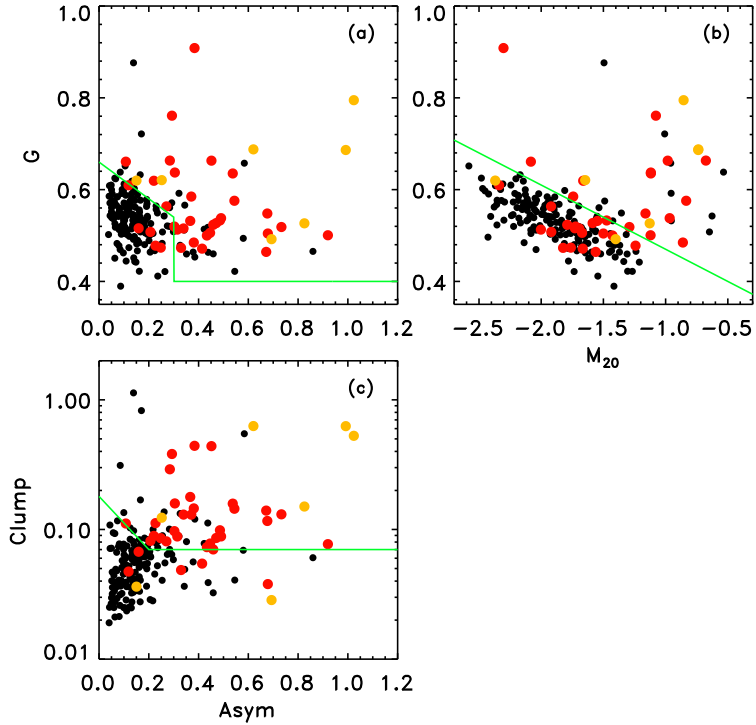}
\caption{ Different diagnostic plots for morphologically disturbed galaxies based on the four parameters {\it Clump}, {\it Asym}, $M_{20}$ and the Gini index.  In panel (b), the green line is the cut proposed by \citet{lotz08} to identify major mergers.  Plotted are all galaxies detected in CO.  The colored symbols represent the galaxies visually identified as being interacting/merging.  The red symbols are the galaxies disturbed morphologically (with or without close companions), and the orange symbols represent the major mergers.  \label{morphparams}}
\end{figure}

In Figure \ref{morphparams} the COLD GASS galaxies with a detection of the CO line are shown in different diagnostic plots that combine the four morphological parameters ({\it Asym, Clump, G} and $M_{20}$).   The $M_{20}-G$ plane was used by \citet{lotz08} to identify major mergers.  Figure \ref{morphparams}b shows that their criterion ($G>-0.14M_{20}+0.33$) does indeed pick up the most extreme systems, but not the full range of morphologies we wish to track.  

The visually-identified disturbed galaxies however cleanly separate from the rest of the sample in the {\it Asym-G} and {\it Asym-Clump} planes (Figs. \ref{morphparams}a,c).  The {\it Asym} parameter appears to be the most sensitive tracer of lopsidedness and low surface brightness tidal features in the outer disks.  A threshold of $Asym>0.35$ was for example suggested by \citet{conselice03} to identify merging galaxies.  There are a number of outliers that are not visually identified as disturbed, but located in Figure \ref{morphparams} above the various selection thresholds.  These galaxies tend to have the smallest angular sizes ($r_{50,r}<5\arcsec$) or to have high inclinations ($i>60^{\circ}$), two situations where morphological classifications, both automated and visual, tend to be less reliable. 

\begin{figure}
\epsscale{1.1}
\plotone{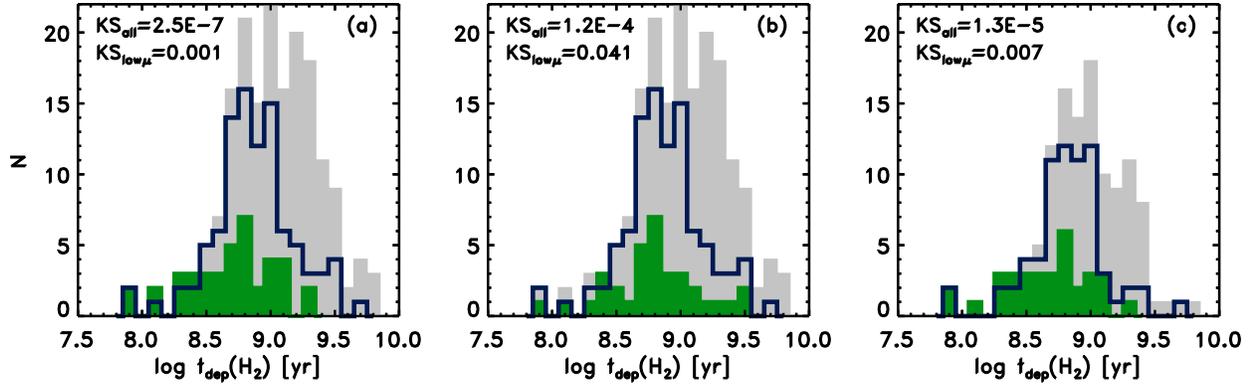}
\caption{Molecular gas depletion time distributions for various subsamples among the COLD GASS galaxies with a secure detection of the CO line ($S/N>5$).  In each panel, the light gray histogram shows the parent population considered in this specific case, the blue solid line the galaxies within this parent population having $log \mu_{\ast}<8.7$\msun kpc$^{-2}$, and the filled dark green the sub-population classified as disturbed/interacting.  The specific selection criteria in the three cases are as follows.  (a) Disturbed galaxies are identified based on the visual classification, (b) Disturbed galaxies are selected in the $Asym-Clump$ plane (Fig. \ref{morphparams}c), (c) The parent sample is restricted to galaxies with inclinations $<60^{\circ}$, to avoid the most inclined galaxies where morphological classifications are more uncertain, and the disturbed galaxies are identified based on the visual classification.  In each case, the KS probability that samples are drawn from the same parent population are given for two sets of samples: KS$_{all}$ compares the parent sample and the disturbed subsample (i.e. the gray and green filled histograms, respectively).  {\em No matter the selection criteria, the depletion times of the disturbed subsample is always significantly different compare to the entire gas-rich population}.  The second value, KS$_{low\mu}$, compares the disturbed subsample among the population below the critical value of $log \mu_{\ast}=8.7$\msun kpc$^{-2}$ (i.e. the greed filled histogram and the open blue histogram).   \label{testsclass}}
\end{figure}

The analysis presented in \S \ref{highSFE} is based on the visual morphological classification.  To confirm that our observation of shorter molecular gas depletion times in morphologically disturbed galaxies is not simply the product of our classification scheme, we repeat the analysis shown in Figure \ref{tdephists} on the full sample with different selection criteria.  The results are shown in Figure \ref{testsclass}.  We compare the initial result (panel a) with the cases when either the morphological parameters are used to select disturbed galaxies (panel b), or when we reject from the analysis the most edge-on galaxies, where classifications are uncertain (panel c).  In all cases, a KS test confirms with very high significance that \tdep\ is shorter in disturbed galaxies compared to the rest of the gas-rich population.  Even when restricting to the subsample with low stellar mass surface densities, no matter how we define morphologically disturbed galaxies or whether or not we include edge-on galaxies in the analysis, we always find a KS probability $<5\%$ that disturbed and un-disturbed galaxies have similar \tdep\ distributions.  In other words, we confirm that low stellar mass surface densities are linked with lower than average \tdep, and that within this population, disturbed/interacting galaxies have the shortest depletion times.


\section{B.  The choice of \xco\ and its impact on this study}
\label{appendix2}

The total molecular gas mass is commonly derived from the measured CO line luminosity using a conversion factor \xco\ as  $M_{H_2}=\alpha_{CO} L^{\prime}_{CO(1-0)}$, where $L^{\prime}_{CO(1-0)}$ is in (K \kms\ pc$^{2}$).  The exact value of \xco\ and its possible dependence on a range of parameters including metallicity and density is being actively investigated, both with observations \citep[e.g.][]{israel97,dame01,rosolowsky03,blitz07,leroy11, genzel12} and simulations \citep[e.g.][]{glover11,shetty11,narayanan12,feldmann12}. 

There is a general consensus that within the Milky Way \xco$=4.35$ \msun (K \kms\ pc$^{2}$)$^{-1}$, with little spatial variations \citep[e.g.][]{strong96,dame01,abdo10}. There is also evidence for at least two sets of circumstances where \xco\ departs significantly from the Galactic value:  low metallicity environments and the very dense central regions of merging systems.  

We investigated in \citet{COLDGASS2} the possibility that depletion time variations observed in the COLD GASS sample are caused by metallicity-induced changes in \xco.  While a few individual galaxies with \mstar$\sim10^{10}$\msun\ may have a higher-than-galactic \xco, we rule out the possibility that the observed \tdep\ variations across the sample are caused by such \xco\ variations.  The various prescriptions for the metallicity dependence of \xco\ indicate that departures from the Galactic value become measurable at $Z<0.5Z_{\odot}$ (see e.g. \citet{leroy11}, but also \citet{genzel12} for a stronger $Z$-dependence, albeit at $z>1$), while most COLD GASS galaxies have $Z>0.8Z_{\odot}$ on the O3N2 scale of \citet{pettini04}, adopting the solar abundance of \citet{asplund04}.  

\begin{figure}
\epsscale{0.9}
\plotone{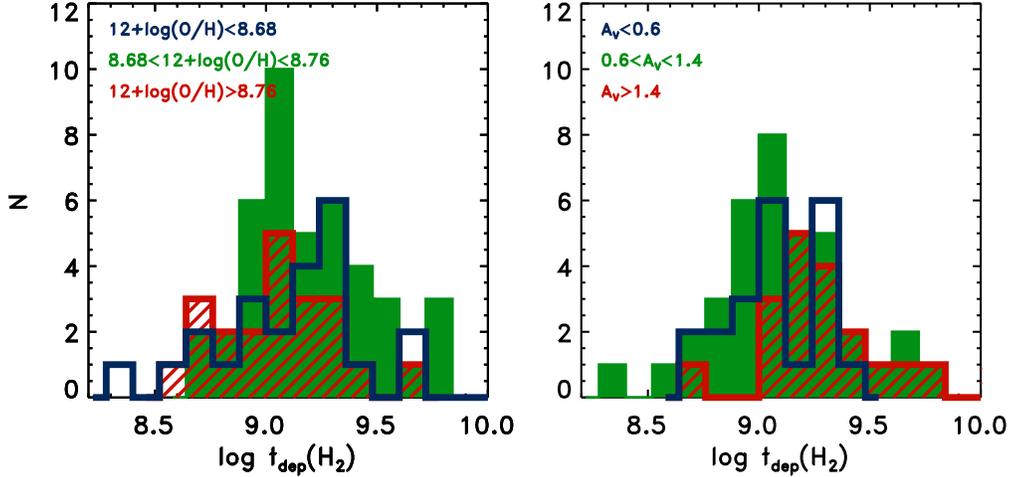}
\caption{{\it Left:} Molecular gas depletion time in three different metallicity intervals: $12+\log({\rm O/H})<8.65$ (open red histogram), $8.65<12+\log({\rm O/H})<8.75$ (dashed blue histogram) and $12+\log({\rm O/H})>8.75$ (filled green histogram).  A KS test indicates that the three distributions are consistent with being drawn form the same parent sample, arguing that metallicity effects on \xco\ are not at the root of the observed depletion time variations.  {\it Right:} Same as left panel, but for three bins of extinction, $A_V$, as measured from the Balmer decrement.  The metallicity and extinction bins are defined such as to isolate the top and bottom quartile in each parameter, respectively.   \label{histZ}}
\end{figure}

For a subset of 196 COLD GASS galaxies, high quality nebular abundances are available from a companion optical spectroscopy observing program \citep{moran12}.  In Figure \ref{histZ}, the depletion time distributions are presented in three intervals of both luminosity-weighted, slit-integrated metallicity and extinction.  The \tdep\ distributions for the three metallicity intervals are statistically equivalent, as confirmed by a KS test.  This is to be compared with for example Figure 10 of \citet{schruba11}; in our stellar mass range (where metallicities are high and their gradients flat), the {\it global} values of \tdep\ do not depend on metallicity.   

Models however suggest that an even better tracer of \xco\ is extinction, as this quantity is the convolution of the metallicity and the volume density \citep{glover11}.   We find that the galaxies with the highest extinctions ($A_V>1.5$) have longer \tdep\ than the rest of the sample (KS$_{prob}=0.04$ that the distributions are drawn from the same parent population), however since these galaxies also have larger than average \must, this may only be the reflection of the previously reported dependence of \tdep\ on \must.  More interestingly, there is no sign for the least extincted galaxies to have shorter than average \tdep; galaxies in the lowest quartile of $A_V$ have \tdep\ values consistent with the rest of the population (KS$_{prob}$=0.56)

\begin{figure}
\epsscale{0.7}
\plotone{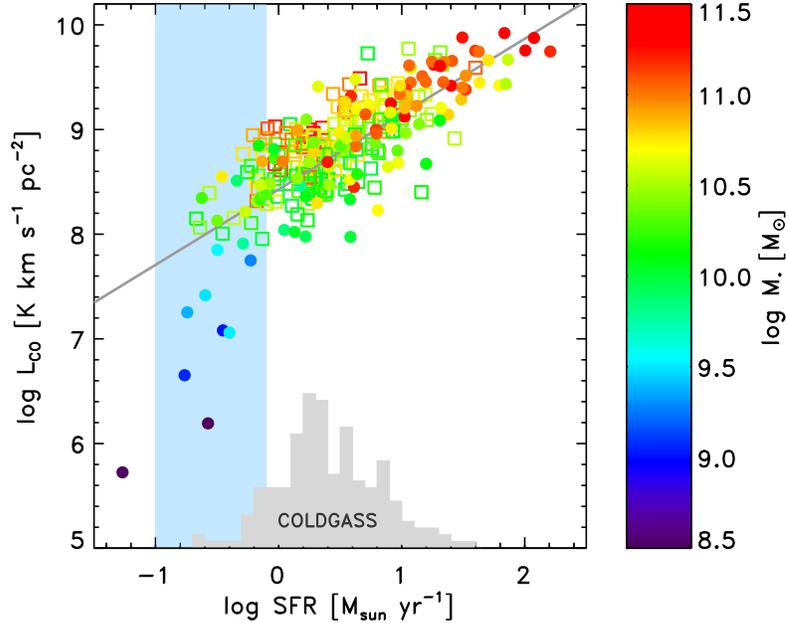}
\caption{Relation between CO(1-0) luminosity and star formation for the COLD GASS sample (open squares) and a compilation of galaxies from the literature (filled circles).  Symbols are color-coded according to their stellar mass.  The break in the relation between $L_{CO}$ and SFR, highlighted by the light-blue shaded region, occurs at \mstar$\sim10^{9.5}$\msun, below the mass range of the COLD GASS survey (\mstar$>10^{10}$\msun).   The histogram shows the SFR distribution of the COLD GASS galaxies with a reliably measured CO(1-0) flux.  \label{LCOSFR}}
\end{figure}

Additionally, as shown in Figure \ref{LCOSFR}, the break in the $L_{CO}$-SFR (or $L_{CO}-L_{FIR}$) relation occurs at \mstar$\sim10^{9.5}$\msun, below the mass range of the COLD GASS sample.  To extend the sample, in particular in the mass range of \mstar$<10^{10}$\msun, which is not probed by COLD GASS, Figure \ref{LCOSFR} also presents data from the literature compilation of \citet{gracia11}, which includes CO measurements from \citet{sanders85,mirabel90,sanders91,combes94,young95,elfhag96,maiolino97,helfer03,gao04,evans05,albrecht04,albrecht07,kuno07,leroy09,young11,garcia12}.   As we further do not observe any dependence of the residuals of the \tdep-\mstar\ relation on metallicity \citep[see Fig. B1 of][]{COLDGASS2}, we conclude that the COLD GASS galaxies, bar a few possible exceptions, are not in the regime where \xco\ is affected by photodissociation of the CO molecules due to reduced dust-shielding. 

\begin{figure}
\epsscale{1.1}
\plotone{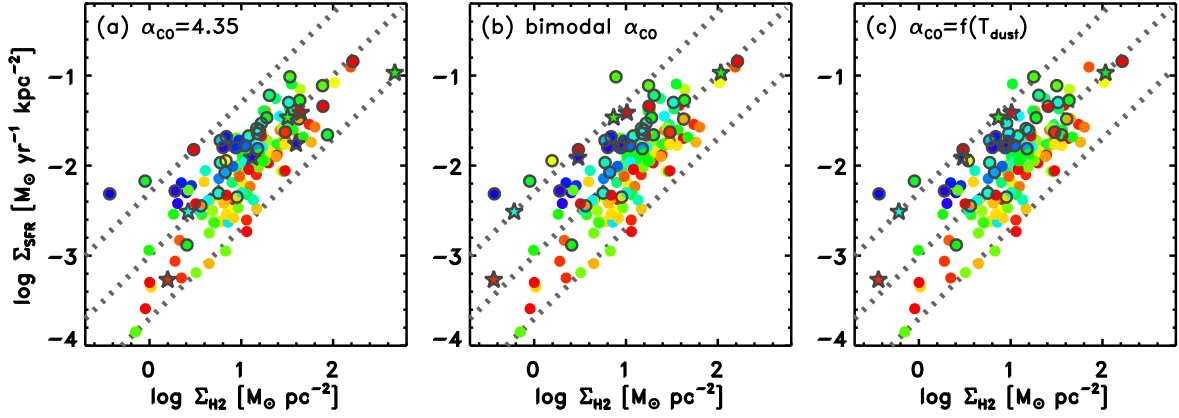}
\caption{The star formation (Kennicutt-Schmidt) relation from the COLD GASS sample for different \xco\ predictions: (a) a Galactic value of \xco$=4.35$ \msun (K \kms\ pc$^{2}$)$^{-1}$ is applied to all galaxies, (b) galaxies are assigned either a Galactic \xco\ or the ``ULIRG" value of 1.0 \msun (K \kms\ pc$^{2}$)$^{-1}$ as explained in \S \ref{gasmass}.  This is the approach used in this paper.  Finally, in panel (c) we use a prescription for \xco\ that depends smoothly on $T_{dust}$ as parametrized by the $S_{60}/S_{100}$ ration \citep{gracia11}. \label{KSXCO}}
\end{figure}

However, some of the galaxies in the sample are interacting, and it has been shown that in systems undergoing major mergers such as nearby ULIRGs and high$-z$ SMGs, the value of \xco\ is reduced by a factor of 5 \citep{solomon97,downes98,tacconi08}. In \S \ref{gasmass} we have presented our favored strategy to assign a value of \xco\ to each galaxy.  In short, a galaxy is given either the Galactic or the ``ULIRG" \xco\ based on a quantitative measure of the ISM conditions, in this case the dust temperature as parametrized by the $S_{60}/S_{100}$ ratio.    As a reference, the KS relation derived with this conversion factor choice and initially presented in Figure \ref{KS} is repeated in Figure \ref{KSXCO}b.  In panel (a) of the same figure, we show the KS relation, had we assigned a Galactic \xco\ for all the galaxies.  The main result of this study, namely that interacting/disturbed galaxies have shorter than average depletion times while bulge-dominated galaxies have the longest depletion times, is still visible.  However, in this case the galaxies undergoing the most violent interactions (and also having the highest IR luminosities and dust temperatures) do no longer mark the upper envelope of the distribution.  The second consequence of applying the Galactic \xco\ to the merging galaxies is that their gas fractions become very large ($0.4<M_{H2}/M{\ast}<1.0$), unrealistically so considering that normal star-forming galaxies in our sample have on average $M_{H2}/M{\ast}=0.09$, and never more that $M_{H2}/M{\ast}=0.3$.  

In Figure \ref{KSXCO}c, we show the results of applying a prescription for \xco\ that also depends on $S_{60}/S_{100}$, but this time in a continuous manner rather than as a bimodal one \citep{gracia11}.  The results are both qualitatively and quantitatively similar.

\end{document}